\documentclass[twocolumn,showpacs,prb,preprintnumbers,amsmath,amssymb]{revtex4}

\usepackage{graphicx}
\usepackage{dcolumn}
\usepackage{bm}

\begin{document}

\preprint{Rapid Communications}

\title{Evidence for multiple impurity bands in sodium-doped silicon MOSFETs}

\author{T.~Ferrus}
\email{taf25@cam.ac.uk}
\author{R.~George}
\author{C.~H.~W.~Barnes}
\author{N.~Lumpkin}
\author{D.~J.~Paul}
\author{M.~Pepper}

\affiliation {Cavendish Laboratory, University of Cambridge, Madingley Road, CB3 0HE, Cambridge, United Kingdom}

\date{\today}

\begin{abstract}

We report measurements of the temperature dependent conductivity in a silicon MOSFET that
contains sodium impurities in the oxide layer. We explain the variation of conductivity in
terms of Coulomb interactions that are partially screened by the proximity of the metal gate. 
The study of the conductivity exponential prefactor and the localisation length as a function of gate 
voltage have allowed us to determine the electronic density of states and has provided arguments
for the presence of two distinct bands and a soft gap at low temperature.

\end{abstract}


\maketitle

Since the invention of the silicon MOSFET, understanding the influence of impurities, especially
sodium contamination, on device performance has been a priority and continues to provide a rich 
system for investigation by experimental and theoretical physicists alike. The electronic
properties of sodium doped MOSFETs were first studied by Fowler and Hartstein \cite {Fowler1, Fowler2}
in the 1970s. They reported a single, broad peak in the subthreshold drain current against gate 
voltage and attributed it to the formation of an impurity band induced by the presence of sodium ions 
near the Si-SiO$_2$ interface. Further studies of narrow channel devices ($\sim$\,100\,nm) demonstrated
a series of reproducible sharp peaks \cite{Fowler3}, while later experiments found evidence for resonant
tunneling between localised states in the channel. \cite{Fowler4, Kopley, Popovic} For sufficiently
low impurity concentrations, the overlap between neighbouring localized electron wavefunctions
and consequently the hybridisation of their excited states is predicted to be reduced \cite{Erginsoy},
splitting the single impurity band observed at high concentrations into the ground and excited bands as
modeled by Ghazali. \cite{Ghazali} Increasing the resistivity of the silicon substrate reduces the scattering
from acceptors at the Si-SiO$_2$ interface, allowing the possibility for such a band splitting to be 
experimentally observed in the transport. In this paper, we will present evidence for the observation 
of two separate impurity bands with a soft gap, based on analysis of the temperature dependent
conductivity below 20\,K.
\newline\indent
The device we used is a MOSFET fabricated on a (100) oriented p-silicon wafer and was subsequently 
patterned in the circular Corbino geometry to eliminate Hall voltages and possible leakage paths. The 
effective gate channel length and interior width were respectively 1\,$\mu$m and 346\,$\mu$m. A high 
resistivity wafer (10$^4$\,$\Omega$.cm) provided a background concentration of less than 
$10^{12}$\,cm$^{-3}$ of boron corresponding to a mean distance between impurities of 1\,$\mu$m. A 
35\,nm gate oxide was grown at 950\,$^{\circ}$C in a dry, chlorine-free oxygen atmosphere. The 
phosphorous implanted and aluminium sputtered contacts were highly metallic and Ohmic at all 
temperatures investigated. Sodium ions were introduced onto the oxide surface by immersing the
device in a $10^{-7}$\,N solution of high purity sodium chloride (99.999\,$\%$) in de-ionised water. The
surface of the chip was dried with nitrogen gas and an aluminium gate subsequently evaporated.
To observe or remove the low temperature conductivity structures, the mobile ions are drifted through 
the oxide to the Si-SiO$_2$ interface, or returned to the Al-SiO$_2$ interface by applying either a $+4$\,V 
or a $-4$\,V DC gate-substrate bias for 10\,min at $65^\circ$C before the device is cooled down to helium
temperature at which sodium looses its diffusivity in the oxide. All measurements were performed using 
standard low-noise lockin techniques with an amplifier of 10$^8$V/A. The AC excitation was maintained 
at 15\,$\mu$V with a frequency of 11\,Hz. Suitable RC filters were employed to eliminate any DC offset from the
amplifier. The gate voltage was controlled by a high resolution digital to analog converter. All experiments
were performed in an $^3$He cryostat and the temperature was measured by a calibrated germanium
thermometer.
\newline\indent
Fig.\,1 shows the conductivity $\sigma$ of our device at 300\,mK versus gate voltage $V_\textup{g}$ for the
case where the sodium ions had been drifted to the Si-SiO$_2$ interface. Two groups of peaks appear clustered
around $V_g\,=\,-2$\,V and $-0.5$\,V and separated by a region of low conductivity and limited by noise. 
The origin of the peaks themselves will be discussed later. Following a $-4$\,V drift, no structure was detectable
over the full range of gate voltages investigated but a difference in threshold voltage of 0.2\,V was found 
at 77\,K between the characteristics of the device following $+\,4$\,V and $-4$\,V drifts. This was attributed
to the presence of mobile charges close to the Si-SiO$_2$ interface at a concentration of 
3.7$\times 10^{11}$\,cm$^{-2}$ corresponding to a mean impurity separation of 16\,$\pm$\,1\,nm. In a reference
device where no sodium was introduced, no sub-threshold conductivity peaks and no shift of the threshold
voltage appeared for any drift conditions investigated. Tunneling through the oxide and other leakage currents
were discounted as the gate leakage current was below 50\,fA at 4.2\,K and was approximately constant over 
the range of gate voltages used. We remark also that no hysteresis and thus no charging effects that have
been reported in similar devices \cite {Travlos} were observed here.
\newline\indent
The presence of two distinct ranges of $V_\textup{g}$ where peaks appear suggests the possibility
of a split impurity band (Fig.\,1). Such a splitting into a ground and an excited state is expected to 
happen for low-doping concentration \cite{Ghazali, Serre} if one takes into account the overlaps 
between impurity wavefunctions and uses a multi-band formalism \cite{Klauder}. The 
conductivity of a Si-MOSFET is not directly related to the density of states and the fact that we see
two regions of high conductivity separated by a region of low conductivity is only an indirect indication
that the device density of states consists of two bands separated by a gap. In fact, through
the Kubo formalism, conductivity tends to be related to local paths through a disordered device but
density of states is a global property.  In order to show that  the density of states splits into two bands,
we have looked at the temperature dependence of the conductivity at a series of different gate voltages.
For all gate voltages studied, the conductivity decreases non-monotonically as temperature is lowered
(Fig.\,2). In the range 1\,K to 20\,K, we observe the characteristics of hopping conduction so that the 
conductivity $\sigma(T)$ is fitted to the generalised equation :

\begin{figure}
\centering
\resizebox{!}{6cm}{\includegraphics{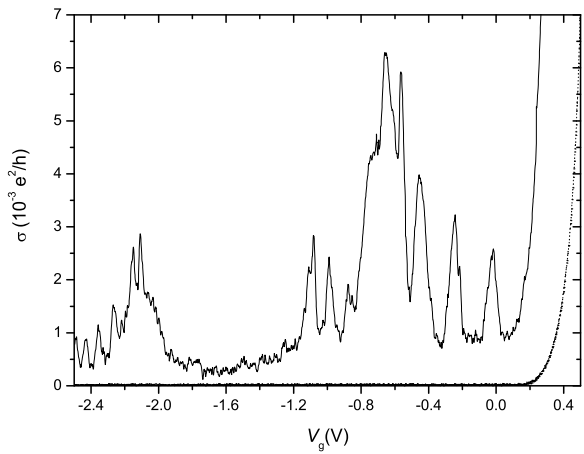}}
\caption{The source drain conductivity versus gate voltage at 300\,mK following a +4\,V drift and a
-4\,V drift (dotted line).}
\end{figure}

\begin{eqnarray}\label{eqn:equation1}
\sigma = \sigma_0\,T^{-\gamma s}\,\textup{e}^{-\left(\frac{T_0}{T}\right)^s}
\end{eqnarray}
\noindent
where $T_0$ and $\sigma_0$ depend on gate voltage. 

\begin{figure}
\centering
\resizebox{!}{5.7cm}{\includegraphics{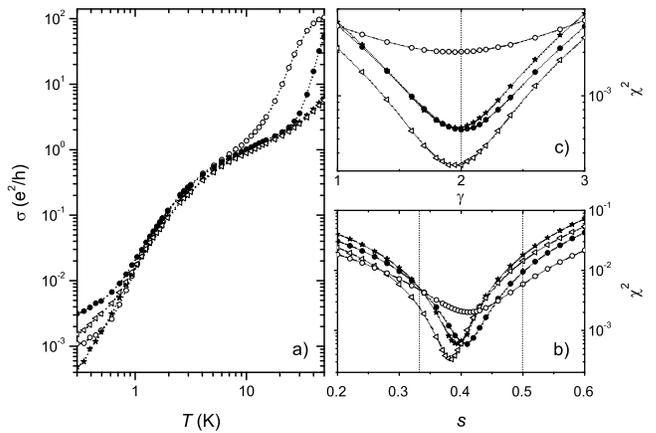}}
\caption{a) Temperature dependence of the conductivity for $V_\textup{g}$\,=\,0.1
($\circ$), $-0.24$ ($\bullet$),$ -1.48$\,V ($\star$) and $-2.26$\, ($\triangleleft$), b) Variation of the 
reduced $\chi^2$ with $s$ for $\gamma\,=\,2$  for the gate voltages listed showing minima at 
$s$\,=\,0.412, 0.406, 0.394 and 0.385 respectively and c) Variation of the reduced $\chi^2$ with 
$\gamma$ for $s$ equal to the optimum value found in 2(b), for the appropriate gate voltages, 
displaying consistency with a minimum at $\gamma\,=\,2$ (dotted line). Lines in b) represent the 
exponent for the Mott hopping regime (left) and Efros and Schklovskii regime (right).}
\end{figure}

\begin{figure}
\centering
\resizebox{!}{6cm}{\includegraphics{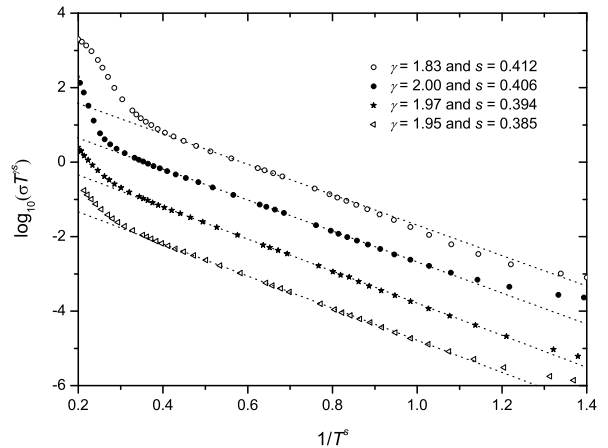}}
\caption{Temperature dependence of the conductivity for $V_\textup{g}$\,=\,0.1 ($\circ$),
$-0.24$ ($\bullet$), $-1.48$\,V ($\star$) and $-2.26$\, ($\triangleleft$) for the optimum values of $\gamma$
and $s$ defined in Eq.\,1. For clarity, curves are shifted downward respectively by 0, 1, 2 and 3 from their 
original values.}
\end{figure}

The best values for $\gamma$ and $s$ are found by minimizing the value of the reduced chi-square 
deviation $\chi^2$ with standard procedures. Figs\,2(a)$-$(b) show the optimum values determined for 
four  gate voltages, one point in the threshold region ($V_g$\,=\,0.1\,V), the upper band ($V_g\,=\,-0.24$\,V), 
the gap ($V_g\,=\,-1.48$\,V) and the lower band ($V_g\,=\,-2.26$\,V). The resulting fits for $\sigma(T)$
are valid over three orders of magnitude in $\sigma$ (Fig.\,3). Studies in gate voltage show that 
$\gamma\,=\,1.98\,\pm\,0.04$ and $s\,=\,0.39\,\pm\,0.02$ for V$_{\textup{g}}$ below 0.25\,V. Above
this point, the value of $s$ decreases rapidly towards $s\,=\,1/3$ for $T\,\geq\,4$\,K and the range is 
better described by Mott hopping conduction \cite{Mott}. It is worth noticing that the smallest values of 
$s$ are obtained for band centre regions $-0.8\,V\,<\,V_{\textup{g}}\,<\,-0.5\,V$ and 
$-2.2\,V\,<\,V_{\textup{g}}\,<\,-2.1\,V$ for which the hopping lengths are relatively smaller and the Coulomb
interactions weaker.
\newline\indent
We emphasise that the use of temperature dependent exponential prefactors allows the fine distinction 
between Mott ($s\,=\,1/3$), Efros-Schklovskii ($s\,=\,1/2$) \cite {Efros} and~the regime under study 
($s\,\sim\,0.39$). The finding that $\gamma\, \sim\,2$ is consistent with the formulation for $\sigma(T)$ 
given by Allen and Adkins \cite{Allen, Mansfield} if it is rederived for the 2D case. This gives 
the conductivity in terms of both the localization length $\xi$ and the density of states at the Fermi level
$n(E_F)$. Accounting for the fact that Coulomb interactions between electrons in different localized states
modifies the density of states close to the Fermi level so that $n(E)\,=\,N_0\,\left|\,E-E_F\,\right|^p$, we
obtain :

\begin{eqnarray}\label{eqn:equation2}
\sigma = \sigma_0\,T^{-\gamma\left(\frac{p+1}{p+3}\right)}\,\textup{e}^{-\left(\frac{T_0}{T}\right)^{\frac{p+1}{p+3}}}
\mbox{ with } \gamma\,=\,2
\end{eqnarray}

\begin{eqnarray}\label{eqn:equation3}
\sigma_0 = \frac{A_0}{\xi^2 \left(p+3\right)^2} {T_0}^{2\left(\frac{p+1}{p+3}\right)}
\end{eqnarray}

\begin{eqnarray}\label{eqn:equation4}
k_BT_0 ={ \left(\frac{p+3}{p+1} \right) } ^ {\frac{p+3}{p+1}}  { \left[ \frac{ \left( {p+1} \right)^3 } 
{ \pi N_0 \xi^2 } \right] }^ {\frac{1}{p+1}}
\end{eqnarray}

\noindent
where $A_0$ is a constant depending on the electronic properties of bulk silicon and $k_B$ the Boltzmann
constant.
\newline\indent
The conditions given by Allen \cite{Allen} for the use of the equations Eqs.\,2-4 are satisfied in our device. 
Recent calculations showed that sodium ions in the oxide may trap either one or two valence electrons 
against the Si-SiO$_2$ interface and that the wavefunctions of the localized states remain hydrogen-like.
\cite{Barnes}  Also, hopping conduction is present in our device but Coulomb interactions are such as the 
only mobile electrons are found in an energy band of few $k_BT$ around the Fermi level. The resulting Coulomb 
gap for $s\,=\,0.39$ (i.e. $n(E)\,\sim\,|\,E-E_F\,|^{0.30}$) is much sharper than for the Efros regime where
$n(E)\,\sim\,|\,E-E_F\,|$. This behaviour has been predicted by Blanter and Raikh \cite{Blanter} while studying
2D systems localized by disorder. They have shown that a metallic gate close to the interface
provides image potentials that modify the density of states to the form $n(E)\,=\,n(E_F)+N_0\,|\,E-E_F\,|^{1/3}$ 
at $T$\,=\,0\,K and thus gives an exponential dependence of $T^{-0.4}$ for the conductivity.  This behaviour is
explained by the fact that the oxide thickness plays the role of the screening length and that initial and final states
become electrostatically independent at low temperature when the hopping length $R$ becomes greater than
twice the oxide thickness $d$. This then produces a crossover from the Efros to the Mott regime. The localization
length in our device was approximately estimated by fitting the conductivity with $p\,=\,0$, using the Mott formula
for $k_B T_0$ and taking the 2D value for the density of states. For $V_g\,=\,-0.4$\,V, this gives 
$\xi\,\sim\,$22\,nm and a hopping length $R\,\sim\,$63\,nm at 1\,K. Our device has a gate oxide of $d\,=\,35$\,nm 
and the Coulomb interactions may be screened by the electrostatic gate as $R\sim\,2d$. This demonstrates the 
device may well be in the regime described by Blanter and Raikh. Fits of the conductivity using $\gamma\,=\,2$
and $s\,=\,0.4$ are still valid for $V_g$ smaller than 0.25\,V and from 1\,K to 18\,K typically, so we will proceed
with these values. In this regime, the density of states at $E_F$ is given by :

\begin{eqnarray}\label{eqn:equation5}
n(E_F)\,=\,n_0(E_F)-\frac{2\pi{n_0(E_F)}^2de^2}{4 \pi \epsilon_0 \epsilon_r} \nu
\end{eqnarray}
\noindent
\begin{eqnarray}\label{eqn:equation6}
n_0(E_F) = {\left( \frac{N_0}{2\pi} \right) }^{\frac{1}{2}}  {\left( \frac{4 \pi \epsilon_0 \epsilon_r}{2 e^2 d^2} \right) }^{\frac{1}{3}}
\end{eqnarray}

\noindent
where $\epsilon_r$ is the permittivity of silicon.
\newline\indent
These equations have been derived at $T\,=\,0$K. At higher temperature but for $k_BT\,\ll\,V(d)$, where $V$ is the coulomb 
potential energy, the gap is partly filled. The Coulomb interactions then become negligible for energies below $V(d)$ and 
the density of states saturates such as $n_0(E_F)\,=\,n_0(E_F+V(d))$.\cite{Aleiner} This condition decreases the value 
of $\nu$ from 1 down to 0.05. Combining (3) and (4), the parameters $N_0$ and $\xi$ were expressed in terms of $T_0$ 
and $\sigma_0$, values easily accessed experimentally. The density of states was extracted using equation (5) and (6).
To find the value of $A_0$, we used $\Xi$\,=\,8.91\,eV for the effective deformation potential of acoustic phonons 
\cite{Fischetti} as well as 3800\,m.s$^{-1}$ for the speed of surface acoustic phonons in silicon \cite{SAW}. This gives 
$A_0$\,=\,6.69 $\times$ $10^{-19}\,e^2.h^{-1}.\textup{m}^2$. 

\begin{figure}
\centering
\resizebox{!}{6cm}{\includegraphics{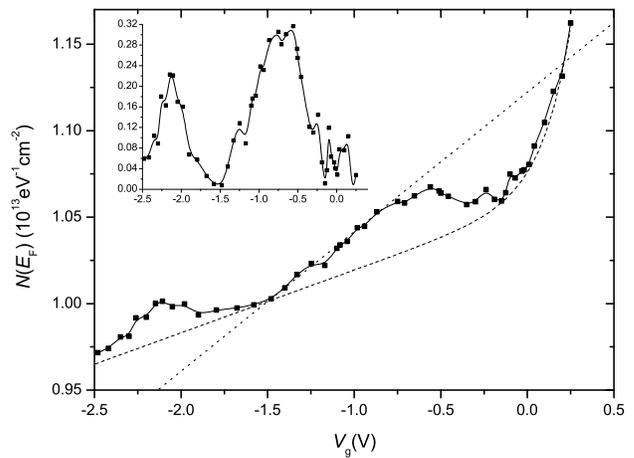}}
\caption{Variation of the DOS at $E_F$ with gate voltage. The dashed and dotted lines represent
respectively the background density due to the conduction band tail and the upper band tail. The inset is the
relative variation of the DOS to the background in the same units.}
\end{figure}

The density of states is shown in Fig.\,4. The value we find is more than one order of magnitude lower than the pure 2D
case ($\sim\,1.6 \times 10^{14}\,$eV$^{-1}.$cm$^{-2}$ with valley degeneracies). The large background is predominantly 
due to the conduction band tail which spreads over the full range of $V_g$ studied. The upper band also has a
significant tail. The presence of density of states tails is a common occurance in disordered systems with low
impurity concentration and localized wavefunctions \cite{Halperin, Zittartz, Kane} but its linear shape for $V_g\,<\,0\,V$
is unusual. It has been attributed to the formation of regions of constant local potential energy at the Si-SiO$_2$
interface and containing a random number of charges.\cite{Arnold} Two regions of higher density are superimposed
on the background density and correspond to the upper and lower groups of peaks. This confirms that the structure
observed in the conductivity is due to the presence of two separate bands. By numerically subtracting the background
density and integrating over the appropriate gate voltage and by supposing a linear relation between the gate voltage
and the surface potential energy, we estimate the upper band contains approximately 3 times the number of states
as the lower band. This value is an upper bound as energetically deeper states could not be accessed experimentally.
\newline\indent
The variation of the localization length $\xi$ (Fig.\,5) follows that of the density of states, showing
the two bands do correspond to the more conductive regions. The value of $\xi$ decreases
rapidly when approaching the threshold voltage. This is expected as the region between
$V_g\,=\,0\,$V and $V_g\,=\,0.4\,$V is both in the band tails of the conduction band and the
upper band, the conduction band edge being well above 0.4\,V. Thus, this region is still
a region of strong localization. Nevertheless, the value of $\xi$ is expected to rise once the
Fermi energy crosses the conduction band edge. Finally, taking into account fitting errors as well as the 
discrepancy in the value of the deformation potential, we estimate the localization length within 5 $\%$ 
and the density of states within 6 $\%$. From the same derivation which gave (1), (2) and (3), the 
transmission coefficient is obtained :

\begin{figure}
\centering
\resizebox{!}{5.9cm}{\includegraphics{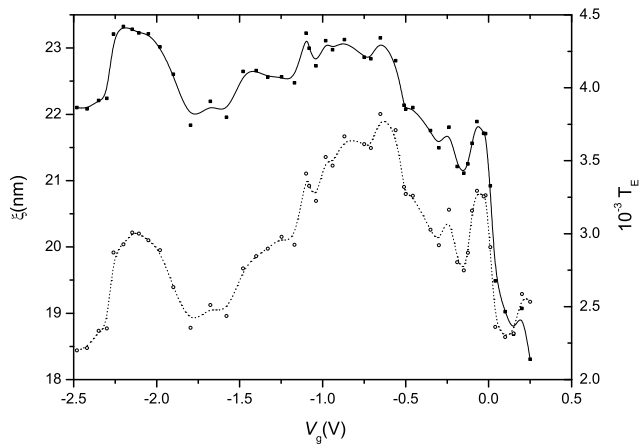}}
\caption{Variation of the localisation length ($\bullet$) and the transmission
coefficient at 1K ($\circ$) with gate voltage.}
\end{figure}
\noindent
\begin{eqnarray}\label{eqn:equation7}
\textup{ln}\left(T_E\right) =\,-\frac{2}{p+3}\left(\frac{T_0}{T}\right)^{\frac{p+1}{p+3}}
\end{eqnarray}

The experimental variation of the transmission coefficient in gate voltage (Fig.\,5)
shows that the conductivity of the two bands mostly comes from the higher mobility
of the states of energies within the upper and lower bands and less from the increase in the
density of states.
\newline\indent
In conclusion, we have observed an unusual hopping regime with an exponent 0.4 which results from the
screening of the Coulomb interactions by the metal gate. We have shown that, consequently, both the 
localization length and the density of states can be extracted from the temperature dependence
of the conductivity. This analysis has given strong evidence for the existence of two separate bands and
a soft gap at low temperature. This may result from the splitting of the impurity band into a lower and
an upper band in presence of Coulomb interactions. The formation of the two bands results itself from the 
presence of a low concentration of sodium impurities close to the Si-SiO$_2$ interface. The conditions
for the observation of such a formation may be the creation of deep but well separated impurity potentials
at the interface, resulting in electron localization. Finally, the electron screening may be sufficiently weak
and the disorder not too strong to allow a Mott-Hubbard transition to take place. The two bands could
then possibly be Hubbard bands.
\newline\indent
We would like to thank Drs T. Bouchet and F. Torregrossa from Ion Beam System-France for
the process in the device as well as funding from the U.S. ARDA through U.S. ARO grant number DAAD19-01-1-0552.

\end{document}